# An Overview on 3GPP Device-to-Device Proximity Services

Xingqin Lin and Jeffrey G. Andrews, The University of Texas at Austin

Amitava Ghosh and Rapeepat Ratasuk, Nokia Solutions and Networks (NSN)

September 26, 2013

Contact Author: jandrews@ece.utexas.edu

## Abstract

Device-to-device (D2D) communication will likely be added to LTE in 3GPP Release 12. In principle, exploiting direct communication between nearby mobile devices will improve spectrum utilization, overall throughput, and energy consumption, while enabling new peer-to-peer and location-based applications and services. D2D-enabled LTE devices can also become competitive for fallback public safety networks, that must function when cellular networks are not available, or fail. Introducing D2D poses many challenges and risks to the long-standing cellular architecture, which is centered around the base station. We provide an overview on D2D standardization activities in 3GPP, identify outstanding technical challenges, draw lessons from initial evaluation studies, and summarize "best practices" in the design of a D2D-enabled air interface for LTE-based cellular networks.

## I. Introduction

Recently, there has been a surge of interest in supporting direct device-to-device (D2D) communication. This new interest is motivated by several factors, including the popularity of proximity-based services, driven largely by social networking applications; the crushing data demands on cellular spectrum, much of which is localized traffic; and the under-utilization of uplink frequency bands. Efforts have been taken by wireless engineers to meet this socio-technological trend: Qualcomm pioneered a mobile communication system known as FlashLinQ wherein "wireless sense" is implemented to enable proximity-aware communication among devices [1]. Moreover, 3GPP is targeting the availability of D2D communication in LTE Release 12 to enable LTE become a competitive broadband communication technology for public safety networks [5], used by first responders. Due to the legacy issues and budget constraints, current public safety networks are still mainly based on obsolete 2G technologies like Project 25 (P25) and Terrestrial Trunked Radio (TETRA) [14] while commercial networks are rapidly migrating to LTE. This evolution gap and the desire for enhanced services have led to global attempts to upgrade existing public safety networks. For example, the USA has decided to build an LTE-based public safety network in the 700 MHz band. Compared to commercial networks, public safety networks have much more stringent service requirements (e.g. reliability and security) and also require direct communication, especially when cellular coverage fails or is not available [5] [14]. This essential direct mode feature is currently missing in LTE.

From a technical perspective, exploiting the nature proximity of communicating devices may provide multiple performance benefits [6]. First, D2D user equipments (UEs) may enjoy high data rate and low end-to-end delay due to the short-range direct communication. Second, it is more resource-efficient for proximate UEs to communicate directly with each other, versus routing through an Evolved Node B (eNB) and possibly the core network. In particular, compared to normal downlink/uplink cellular communication, direct communication saves energy and improves radio resource utilization. Third, switching from an infrastructure path to a direct path offloads cellular traffic, alleviating congestion, and thus benefitting other non-D2D UEs as well. Other benefits may be envisioned such as range extension via UE-to-UE relaying.

From an economic perspective, LTE D2D should create new business opportunities, though its commercial applications are not the focus in LTE Release 12. For example, many social networking applications rely on the ability to discover users that are in proximity, but the device discovery processes (e.g. Facebook Places) typically work in a non-autonomous manner. Users first register their location information in a central server once launching the application; the central server then distributes the registered location information to other users using the application. It would be appealing to the service providers if device discovery can work autonomously without manual location registration. Other examples include e-commerce, whereby private information need only be shared locally between two parties, and large file transfers, e.g. just-taken video clips shared amongst other nearby friends.

Thus far, use cases of 3GPP proximity services (ProSe) have been specified in [2] and the corresponding architecture enhancements are studied in [3]. In addition, a new D2D study item was agreed upon at the December 2012 RAN plenary meeting [4]. Through the most recent 3GPP meetings, initial progress on D2D evaluation methodology and channel models has been made [7], and 3GPP recently agreed that for LTE Release 12, ProSe would focus on public safety networks, especially the one to many communications [15].

## II. Overview of 3GPP Proximity Services (ProSe)

In this section, we provide a brief tutorial on the fundamentals of 3GPP ProSe, including basic use cases, scenarios, objectives, evaluation methodology and channel models. These aspects lay the foundation for the ProSe design.

### Basic Functions and Scenarios

D2D discovery and D2D communication are the two basic functions for supporting 3GPP ProSe services. All the ProSe use cases studied in [2] depend on them. From a UE's perspective, D2D discovery enables it to use the LTE air interface to identify other UEs that are in proximity. D2D discovery may be broadly classified into two categories: restricted discovery and open discovery, in terms of whether permission is needed or not. D2D communication is the communication between two UEs in proximity using LTE air interface to set up a direct link without routing via eNB(s) and possibly core network.[1] Here proximity should be understood in a broader sense than just physical distance. It may be also determined based on e.g. channel conditions, SINR, throughput, delay, density, and load.

---

[1] Note that in [2], D2D communication can also refer the communication between two UEs in proximity whose data are routed via local eNB (but not involving Serving Gateway/Packet Data Network Gateway). In this paper we restrict the concept of D2D communication to direct mode only.

For the ease of evaluation, 3GPP categorizes D2D scenarios in terms of the presence of network coverage. In the in-coverage scenario all the UEs are covered by the eNBs while in the out-of-coverage scenario no UE can be covered by the eNBs.[2] Partial-coverage scenario lies somewhere in between: Some UEs are in coverage while the remaining UEs are not. The evaluation of partial-coverage scenario may be performed by disabling a fraction of eNBs from the in-coverage scenario.

### D2D vs. Ad Hoc Networks

Before delving into the D2D analysis and design, it is helpful to contrast D2D with mobile ad hoc networks (MANET), which have been studied and developed extensively over about 3 decades, with very limited success, for reasons partially documented in [13]. A key difference is that D2D can typically rely on assistance from the network infrastructure, i.e. base stations, for control functions like synchronization, session setup, resource allocation, routing, and other overhead-consuming functions that are extremely costly in a MANET. Further, D2D networking mainly consists of local, opportunistic, and single-hop communication, whereas multi-hop routing is typically needed in a MANET and long hops may be unavoidable, which hurt network performance. In D2D-enabled cellular, we only do direct communication when it is beneficial, with the BS's providing an efficient fallback.

In the public safety context, D2D must function even without BS support, so is more like a MANET. Service in this out-of-cellular-coverage mode is only required to be rudimentary, so is more like a walkie-talkie than a full MANET that may require streaming video. Further, out-of-coverage public safety UEs are often clustered (on the order of at most tens of nodes) and so the clusterhead can act as the *de facto* BS.

Although simpler than a MANET, adding D2D features to LTE still poses many challenges and risks. Cellular networks have existed for several decades and network operators are sure to resist a technology that takes away their control (exercised mostly at the BS). Further, the idea of a UE-UE link is new: all existing cellular technologies including LTE are designed and optimized for eNB-UE links. Further, the D2D design has to take into account its impact on wide area network (WAN) as a whole. To summarize, the following issues have to be carefully studied in order to support D2D in LTE [4].

---

[2] A UE is said to be out of coverage when the average receive SINR from the network is less than -6dB [7].

1) Identify and evaluate techniques to make UEs capable of D2D discovery and communication
2) Identify and evaluate techniques to manage D2D links
3) Evaluate the impact of ProSe services on metrics like UE battery life, existing operator services (e.g. voice calls) and operator resources (e.g. amount of resource used by D2D discovery)

The general principle for the above studies is to reuse existing LTE features as much as possible [4].

## Evaluation Methodology and Channel Models

Each of the aforementioned D2D studies involves rich design/research topics and requires comparison of different technical options. As a starting point, it is necessary to agree on a common evaluation methodology. Much progress has been made in 3GPP on this aspect. Evaluation assumptions (including carrier frequency, system bandwidth, UE density and mobility, RF parameters and traffic models) as well as performance metrics may be found in [7]. Note that there is a continuing debate in 3GPP on network layout for evaluating D2D. As a result, six diverse layout options are available [7]. For example, Options 1 and 4 incorporate remote radio head (RRH) or indoor hotzone while the remaining options consider macrocells only.

Appropriate channel models are important for generating realistic results for D2D evaluation in both link and system simulations. Existing asymmetric eNB-UE channel models are ill suited for modeling the symmetric UE-UE channels. Specifically, the following factors make the propagation characteristics of UE-UE links distinct from those of eNB-UE links [9].

**Dual mobility.** In eNB-UE links, only the UEs are mobile while the eNBs are fixed. In contrast, both terminals may be mobile in UE-UE links, creating a dual mobility scenario.

This dual mobility affects the temporal correlation of shadowing as well as fast fading, e.g., increasing Doppler spread.

**Low antenna height.** The antenna height at the eNBs may range from several meters (for femto eNBs) to tens of meters (for macro eNBs), while the typical antenna height at the UE is 1.5 m. With the same link length, UE-UE link incurs higher pathloss than eNB-UE link. In addition, since both terminals of a UE-UE link are low, they see similar near street scattering environment, which is different from the scattering environment around an eNB.

**Interlink correlation.** It is expected that D2D UEs are of high density; for example, 150 UEs per cell are assumed in 3GPP for evaluating D2D discovery [7]. As a result, small inter-UE distances are expected; and compared to eNB-UE links, there exist much higher correlations in the propagation characteristics of UE-UE links including shadowing, angle of arrival (AoA) and angle of departure (AoD) spreads, delay spread.

Due to the above distinctive traits of UE-UE links, the ideal approach would be to conduct realistic measurements and develop appropriate D2D channel models. However, this may significantly slow down the progress of the D2D study item [4], so the general philosophy adopted by 3GPP is to adapt existing channel models to D2D, summarized in Table 1.

|  | Outdoor to outdoor | Outdoor to indoor | Indoor to indoor |
|---|---|---|---|
| Pathloss | Winner+ B1 with -10 dB offset | Dual strip (for Option 2); Winner+ B4 (for other Options) with -10 dB offset | Dual strip (for Option 2); ITU-R InH (for other Options) with LOS prob. given in ITU-R UMi |
| Shadowing | 7 dB log-normal; assumed i.i.d., i.e., correlation is not modeled yet |||
| Fast fading | Agreed to use symmetric angular (i.e. AoD and AoA) spread distribution |||
|  | Agreed to amend ITU-R UMi/InH[3] to account for dual mobility |||

Table 1. 3GPP TSG RAN WG1 #73 way forward on D2D channel models

---

[3] ITU channel models: indoor hotspot scenario (InH) used for ITU Indoor; urban micro-cell scenario (Umi) used for ITU Microcellular.

# III. Design Aspects

D2D ProSe is a relatively new study item and its design is largely open. In this section, we provide an overview on its design aspects being discussed in 3GPP, and organize them into four broad topics: ProSe management, synchronization, device discovery and direct communication. Design options are compared throughout this section, with a summary in Table 2. The treatment is mainly from a radio access perspective; higher layer issues like security, authorization, privacy and billing may be found in [3].

## ProSe Management

**Control mode (ad hoc vs. clusterhead).** In cellular networks including LTE, the control plane only exists between the UE and network, i.e., the network fully controls the operation of mobiles except standardized and vendor-specific aspects. When D2D UEs are in coverage, ProSe services should be under continuous network management and control; for example, D2D mode selection (i.e. direct path vs. infrastructure path) shall be determined by the network. However, full network control over D2D UE behaviors may be over-designed. For example, allowing hybrid automatic repeat request (HARQ) operation to be directly handled by D2D UEs may alleviate the network burden and reduce feedback delay. Similarly, link adaptation may be directly handled by D2D UEs. These observations motivate the necessity of the split of the control functionality between the network and UE. The specific split requires detailed analysis and study.

Further, D2D UEs may enter out-of-coverage area, in which the network loses its control capability. Two alternative control topologies are shown in Fig. 1: Ad hoc and clusterhead-based. In the ad hoc topology, each D2D UE controls its own behavior and the transmissions may be coordinated based on random MAC protocols like CSMA. This control mode is simple in terms of implementation. However, random MAC protocols are not as efficient as centralized scheduling. Also, they do not fit well into existing LTE architecture and thus significant re-engineering of LTE would be required.

| | Design options | Pros | Cons |
|---|---|---|---|
| **Out-of-coverage control mode** | Master-slave (Clusterhead) | Similar topology as E-UTRAN<br>Reuse existing eNB functions | Requires a clustering scheme<br>The master UE becomes the bottleneck<br>Places large burden on the master UE, who may "revolt" |
| | Ad hoc | Distributed and "democratic" | Deviates a lot from E-UTRAN topology<br>Overhead can be extremely high |
| **Resource allocation for D2D signals** | Static | Low complexity and overhead<br>Suitable for device discovery use case | Cannot adapt to dynamic traffic demands |
| | Dynamic | Resources are used more flexible | High complexity and overhead |
| **Use of radio resources** | Downlink | Can reuse downlink Rx chain | Interference to the downlink reception at UE<br>Downlink resources are congested<br>Heavy control signaling exists in the downlink<br>Regulatory constraints (for FDD LTE)<br>Requires downlink Tx chain at UE (for FDD) |
| | Uplink | Can reuse uplink Tx chain | Interference to the uplink reception at eNB<br>Requires uplink Rx chain at UE (for FDD) |
| **Modulation format** | SC-FDMA | Low PAPR<br>Current UE Tx format | Requires new SC-FDMA receiver at UE |
| | OFDMA | The complexity of adding OFDMA Tx is lower than adding SC-FDMA Rx | High PAPR, leading to lower power efficiency<br>Lower range |
| **HARQ** | Direct | Uniform solution for both in-coverage and out-of-coverage D2D | HARQ feedback is not known to the network |
| | Indirect (routed by the network) | May reuse existing downlink and uplink channels with minimal changes | High overhead<br>Longer feedback delay |
| **Discovery type** | Direct discovery | Does not require cellular coverage<br>Better suited for open discovery | More frequent discovery signal transmissions<br>Impact on UE battery life |
| | EPC-level discovery | Requires least effort from UE<br>Minimizes impact on UE battery life | Requires EPC to track UE at fine resolution<br>Relies on cellular coverage |
| **Discovery signal** | Sequence-based | Low complexity transmission/reception<br>The sequence may be reused for synchronization | Information conveyed is limited |
| | Packet-based | May contain rich information | More complex transmission/reception, e.g., synchronization may be required before decoding |
| **Synchronized/ asynchronous discovery** | Synchronized discovery | Efficient in terms of UE energy and use of radio resources<br>Faster and more reliable (i.e. fewer false alarms) | Synchronization before discovery may not be available in the out-of-coverage scenarios |
| | Asynchronous discovery | Works for both in-coverage and out-of-coverage scenarios | Low efficiency in terms of UE energy and use of radio resources |

Table 2. Comparison of design options for 3GPP ProSe

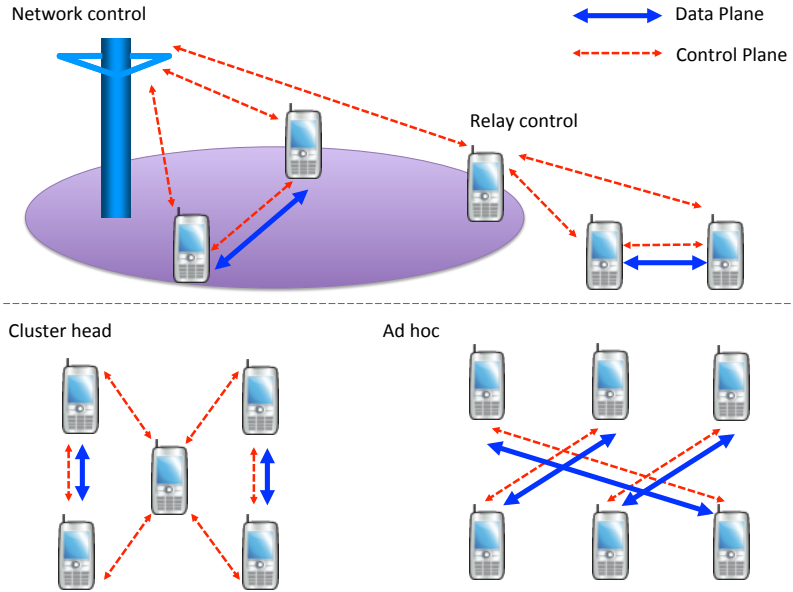

Fig. 1. Control modes for different D2D scenarios

In the clusterhead based control topology, one UE assumes a master role and acts as clusterhead within a group of UEs (see e.g. [8]). The clusterhead is like an eNB and can help achieve local synchronization, manage radio resources, schedule D2D transmissions, etc., for slave UEs in its cluster. This mode makes the out-of-coverage ProSe topology (at least from the control plane standpoint) similar to E-UTRAN where eNB serves UEs in its cell, and has the advantage that many existing functions of E-UTRAN may be applied (possibly with appropriate modification) to out-of-coverage D2D. This may save standardization effort in 3GPP. The disadvantages are that the clusterhead becomes the control bottleneck and its battery gets drained.

Note that the master-slave control mode is not restricted to the out-of-coverage scenario. For example, an authorized in-coverage UE may assume a master role and control out-of-coverage UEs that are in its range. Alternatively, it may act as a relay to receive and re-transmit control signals from the eNB to out-of-coverage UEs, as shown in Fig. 1.

**Downlink vs. Uplink.** While public safety UEs normally have access to dedicated spectrum, commercial D2D UEs have to share the radio resources with existing cellular UEs in either paired frequency-division duplexing (FDD) or unpaired time-division duplexing (TDD) LTE networks. This leads to the question: Which part of the radio resources should D2D transmission utilize, downlink or uplink resources or both? Different choices lead to quite different interference situations. In particular, when D2D transmission utilizes downlink resources, a transmitting D2D UE may cause high interference to nearby cochannel cellular UEs receiving downlink traffic. In contrast, when D2D transmission utilizes uplink resources, the receiving D2D UE experiences strong interference from nearby cochannel cellular UEs transmitting uplink traffic.

There are several good reasons for favoring using uplink resources. First, the uplink resources are often less utilized compared to the downlink resources, and thus sharing the uplink resources with D2D may improve spectrum utilization. Second, downlink channel contains heavy control signaling. To minimize the impact of D2D on WAN performance, complicated design is required if not infeasible. Third, reusing uplink resources can minimize the D2D interference to cellular transmission as the interference can be better dealt by eNBs, which are typically located far away from UEs and more powerful. Last but not least, in FDD LTE, reusing uplink resources requires the UE to be capable of receiving in the uplink while reusing downlink resources requires UE to be capable of transmitting in the downlink. In addition to regulatory concerns, the latter is more complicated in terms of hardware design (due to the more stringent transmit RF requirements).

**Resource management**. When UEs are in coverage, the network is responsible for radio resource management. In principle, the network can allocate the resources either dynamically (i.e. based on current D2D transmission demand) or statically (i.e., certain resources are periodically reserved for D2D transmission). Clearly, dynamic allocation utilizes the radio resources more flexibly at the cost of heavy control overhead while the converse is true for static allocation.

For D2D discovery, static allocation seems appropriate. If radio resources are allocated dynamically, UEs need to be continuously active, which leads to high energy consumption. In contrast, static allocation may minimize the impact of discovery on UE battery. For example, a frame structure may be standardized such that 50 contiguous uplink subframes in every 5 seconds are reserved for discovery, consuming just 1% of the network capacity. This allows the UEs involved in discovery to sleep for 99% of the time and only wake up to transmit/receive discovery signals in the predefined subframes.

For D2D communication, dynamic allocation is more appropriate than static allocation, since the fluctuation of D2D traffic may vary significantly over both space and time. Note that if the D2D UE density is high, centralized resource scheduling on the time scale of 1 ms (i.e. current LTE scheduling time scale) may involve high overhead for collecting the UE-UE channel state information (CSI) and then informing the UEs about the scheduling decisions. Alternatively, the network may simply allocate a resource pool for D2D communication and let the D2D UEs contend for it using random access protocols.

When UEs are out-of-coverage, radio resources may be managed in a centralized manner (e.g. by the clusterhead). Alternatively, UE may be preconfigured with a distributed resource access protocol (e.g. CSMA), which can be launched when the UE enters an out-of-coverage area.

### Synchronization

Synchronized D2D transmissions are appealing, and this is one major advantage of in-coverage D2D networking vs. MANET: The eNB provides a synchronization beacon. For example, with time synchronized device discovery UEs can be active only during the pre-determined time slots for receiving discovery related signals. This consumes much less energy than the asynchronous discovery where continuous searching for discovery signals may be required. However, synchronization for D2D transmissions is challenging because it typically involves multiple D2D links: The signals are emitted from different transmitting UEs (contrary to the downlink situation) and arrive at different receiving UEs (contrary to the uplink situation).

When D2D UEs are in coverage and synchronized to their corresponding eNBs, the first question in FDD LTE is whether to select uplink or downlink timing for D2D transmission. If the uplink band is used for communication, then using uplink timing for D2D transmission may generate less interference. Nevertheless, neither of them may guarantee synchronization between two D2D UEs because (1) they may be associated with different eNBs that are not synchronized in FDD LTE, and (2) even located in the same cell, they may have different distances to the eNB and different timing advance adjustments may be applied. The last issue also exists in TDD LTE. Thus, the impact of timing misalignment on link/system performance deserves further study, and additional synchronization methods are needed if the impact turns out non-negligible.

Synchronization becomes more challenging when UEs are out of coverage. In this case, periodic transmission of synchronization signals from UEs may be needed. Though reuse of existing LTE overhead signals like primary/secondary synchronization signal (PSS/SSS) is simple and obvious, it is not *a priori* clear if they will suffice or further optimization is needed. Further, the design of synchronization signals including the transmission period, radio resource and transmit power is also an open issue. One straightforward solution is to use the clusterhead based control mode, in which case the clusterhead transmits the synchronization reference signal. Besides, an authorized in-coverage UE may send or relay the eNB's synchronization reference signal to out-of-coverage UEs that are in its communication range.

### Device Discovery

The capability of detecting nearby UEs is required for both commercial and public safety UEs [2]. Device discovery may be broadly categorized into two types: Direct discovery and evolved packet core (EPC)-level discovery [3]. In the case of direct discovery, UE would search for nearby UEs autonomously; this requires UEs participating in the device discovery process to periodically transmit/receive discovery signals. Two discovery mechanisms are possible – a push mechanism where UE broadcasts its presence and a pull mechanism where UE requests information regarding discoverable UEs. Direct discovery works in both in-coverage and out-of-coverage scenarios and does not preclude network assistance when available. In the case of EPC-level discovery, EPC determines the proximity of UEs and a UE starts the device discovery process after it receives its target information from the network. This scheme requires the network to keep track of the UEs, reducing the discovery burden on the UEs.

**Discovery signal design.** With either direct or EPC-level discovery, UEs will transmit discovery signals that may be detected by other UEs. A natural question is what kind of information should be carried by the discovery signals. Currently, it is assumed in [3] that UE identity would be included; other contents like application-related information may be included. The amount of information sent during discovery determines the required amount of radio resources and also affects the discovery signal or channel structure. Thus, a rough estimate of the quantity of discovery information may facilitate the design.

If the size of discovery information is small, it may be sufficient for UEs to transmit certain sequences for discovery. Though only limited information may be conveyed, transmission/reception of these sequences is of relatively low complexity. As a starting point, it is natural to evaluate if existing LTE physical layer signals such as PSS/SSS, PRACH and the various types of reference signals suffice. Then a decision on whether to reuse existing signals or design new sequences can be made. Note that these sequences may also be used for synchronization purpose. If the size of discovery information is too large to be handled by sequence-based design, a packet-based design may be used [10]. Though transmission/reception of discovery packets is more complex, robust channel coding may be used to improve the discovery reliability at the cost of additional complexity.

**Synchronous vs. asynchronous discovery.** There is also a debate in 3GPP regarding whether synchronization should be assumed for device discovery. Compared to asynchronous discovery, synchronous schemes are obviously appealing as they are more efficient in terms of energy consumption and spectral efficiency and result in more reliable, faster discovery. However, assuming synchronization *a priori* before device discovery may be questionable in the out-of-coverage scenario. This implies that at least for public safety networks, UEs may need asynchronous discovery capabilities.

## Direct Communication

Several direct communication modes are defined in [4] including unicast, relay, groupcast and broadcast. Though reusing some of the existing LTE design (like the frame structure and frequency parameters) is possible, supporting UE-UE communication may require many physical layer changes and new standardization efforts, as detailed below.

**Modulation format.** The first question on direct communication is on the selection of waveform format. Currently, LTE uses SC-FDMA in the uplink and OFDMA in the downlink, so the UE is equipped with an SC-FDMA transmitter and OFDMA receiver. If SC-FDMA is used (resp. OFDMA), the D2D UE needs to be equipped with a new SC-FDMA receiver (resp. OFDMA transmitter). Compared to implementing OFDMA transmitter, implementing SC-FDMA receiver is more complex since the single carrier transmission requires relatively complex equalization at the receiver. However, SC-FDMA transmitter can enjoy low peak to average power ratio (PAPR).

**Power control.** D2D power control is useful in saving UE energy and reducing interference. Note that uplink transmit power is fully controlled by the eNB. Allowing D2D UE to have some control on its transmit power may reduce control signaling overhead and delay. For example, eNB may just be responsible for open loop power control and setting a coarse transmit power level and permissible power range, while the D2D UEs can handle finer closed loop power control to adapt to fast channel quality variations.

**Channel measurements.** For ProSe management purpose, the network should know the channel condition of D2D links [2], making channel measurement an indispensable component of D2D communication. Depending on the control mode, the measurement results may be reported to the network or the peer UE. To enable channel measurement, the design of reference signals used for UE-UE links requires further study, though initial channel measurement may be performed during the device discovery process by exploiting the discovery signal. As a starting point, the applicability of existing LTE reference signals can be evaluated. Also, it is desirable to categorize D2D communication according to e.g. the UE-UE range and/or mobility. Then reduction of reference signal overhead may be possible for short and low mobility UE-UE links as the channel should have fewer taps and vary slowly.

**HARQ operation.** HARQ combines forward error correction (FEC) and ARQ retransmission. As the interference situation may be quite complex and dynamic for D2D communication, HARQ would make the D2D communication more robust. D2D HARQ may be either indirect or direct [11]. In indirect HARQ, D2D receiver first sends ACK/NACK to the eNB and then the eNB relays ACK/NACK to the D2D transmitter. Indirect HARQ allows reusing existing LTE downlink and uplink channels with minimal changes at the cost of additional overhead and possibly longer feedback delay. In direct HARQ, D2D receiver directly sends ACK/NACK to the D2D transmitter. Direct HARQ may be used in either in-coverage or out-of-coverage scenario.

Note that for LTE Release 12, ProSe would focus on public safety broadcast [15], which typically does not have closed loop feedback. In this case, HARQ operation will not be supported.

## IV. System-Level Performance of D2D

In this section, we present some initial evaluation results to gain insights into the performance aspects of ProSe services. The major simulation assumptions are as follows. (1) Each hexagonal cell consists of three sectors whose antenna pattern is specified in [12]. (2) Assuming all UEs are located outdoor, UE-UE pathloss model is Winner+ B1 with -10 dB offset (cf. Table 1) and UE-eNB pathloss model is specified in [12]. (3) Given the number of cellular UEs, they are uniformly dropped in each sector. The same dropping is applied to transmitting D2D UEs. (4) For each dropped transmitting D2D UE, a peer receiving D2D UE is dropped uniformly in the ball of some radius (termed D2D range below) centered at the transmitting D2D UE. So the receiving D2D UE may or may not be located in the same sector with its peer. (5) Open loop power control is used:

$$P_t = \min(P_{max}, SNR_t + P_{noise} + \alpha \cdot PL),$$

where $P_t$ denotes the transmit power, $P_{max}$ denotes the peak transmit power and equals 23 dBm throughout the simulation, $SNR_t$ is the adjustable SNR target, $P_{noise}$ denotes the noise power, and $\alpha$ is the pathloss compensation factor, and $PL$ denotes the link path loss (with shadowing included). For the *no power control* case, each UE transmits at its maximum power.

To begin with, we consider the public safety scenario and the SINR distributions of D2D links under different power control settings are shown in Fig. 2. If the 10 dropped D2D transmitters per sector are not coordinated and active simultaneously, the left plot of Fig. 2 shows the resulted undesirable SINR distributions: At most 40% D2D links can have greater than -6 dB SINR. This poor SINR performance is due to the fact that D2D UEs are randomly distributed and the resulting near-far problem cannot be effectively dealt by open loop power control (contrary to the uplink case where all the UEs transmit to the same eNB).

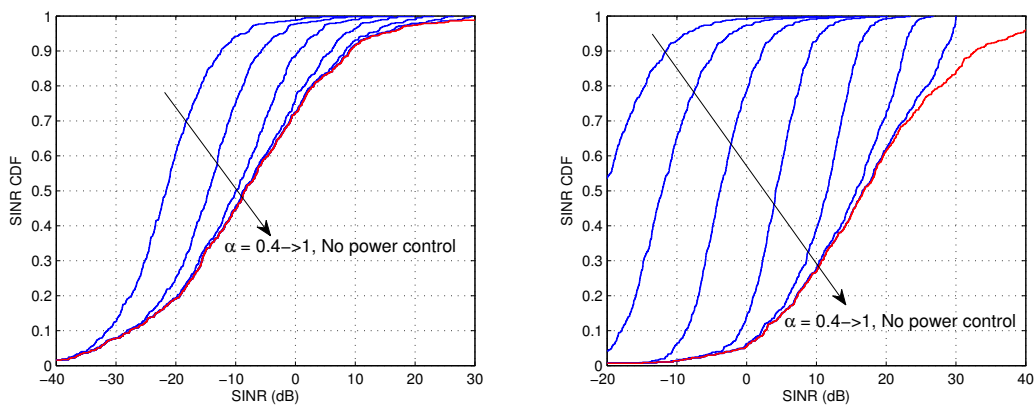

Fig. 2. SINR distribution of D2D links: ISD = 1732 m; D2D range = 250 m; 10 transmitting D2D UEs per sector are active in the left subfigure while 1 transmitting D2D UE per sector is active in the right subfigure.

In contrast, if the 10 D2D links per sector are coordinated and multiplexed orthogonally in the time domain, much better SINR distribution can be obtained, as shown in the right plot of Fig. 2. In this case, with appropriate power control setting more than 95% D2D links can have greater than -6 dB SINR. These simulation results imply that *D2D links have to be coordinated to ensure successful transmissions*. While in-coverage D2D links may be coordinated by eNBs, out-of-coverage D2D UEs have to resort to random MAC protocols or clusterhead based control design.

Next we consider the general scenario and study the impact of D2D range. The SINR distributions of D2D links under different power control settings are shown in Fig. 3. The left plot of Fig. 3 shows that at most 50% D2D links can have greater than -6 dB SINR even with only 1 cochannel transmitting D2D UE per sector. This further implies that supporting long D2D range (important for public safety scenario) requires coordination among eNBs. In contrast, spatial reuse is possible when D2D range is reduced, as shown in the right plot of Fig. 3: With appropriate power control settings, two cochannel transmitting D2D UEs per sector can be supported when the D2D range is 50 m.

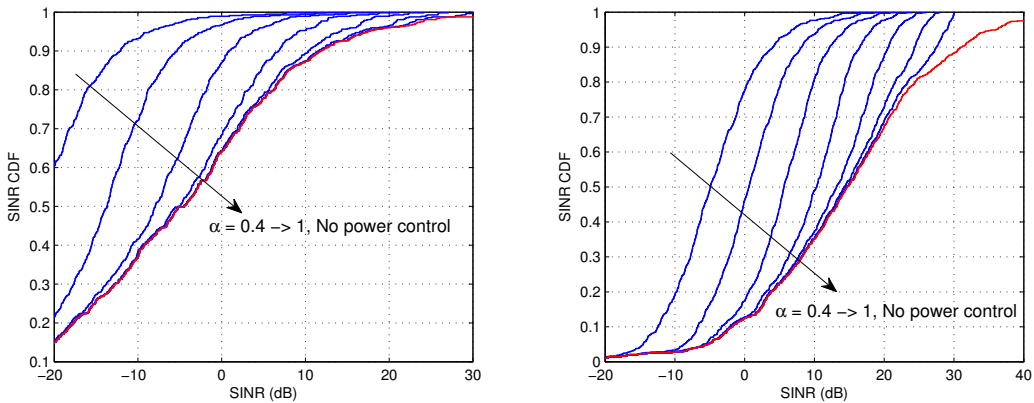

Fig. 3. SINR distribution of D2D links: ISD = 500 m; 1 transmitting D2D UEs/sector are active in the left subfigure with 250 m D2D range while 2 transmitting D2D UEs/sector is active in the right subfigure with 50 m D2D range.

Finally, we evaluate the throughput performance. The upper (resp. bottom) plot of Fig. 4 shows the average throughput (resp. bottom 5% throughput) vs. the number of transmitting D2D UEs. The results show that *offloading by D2D communication can yield throughput gain in terms of both average throughput and bottom 5% throughput.* The gain is more remarkable for the bottom 5% throughput. One interesting phenomenon observed in Fig. 4 is that *the throughput gain decreases when the number of transmitting D2D UEs is large* (e.g. when it equals 9 in Fig. 4). This is because the receiving D2D UEs have more complex interference environment than the eNBs, which becomes more dominant when the number of D2D pairs increases. This offsets the proximity gain.

Note that the presented throughput evaluation is focused on the uplink only; it does not take into account that D2D may further save the downlink and possibly core network resources. Thus the actual throughput gain may be even larger. Besides, more sophisticated scheduling algorithms (e.g. allowing spatial reuse in each sector) may provide additional gains. We treat these as future work.

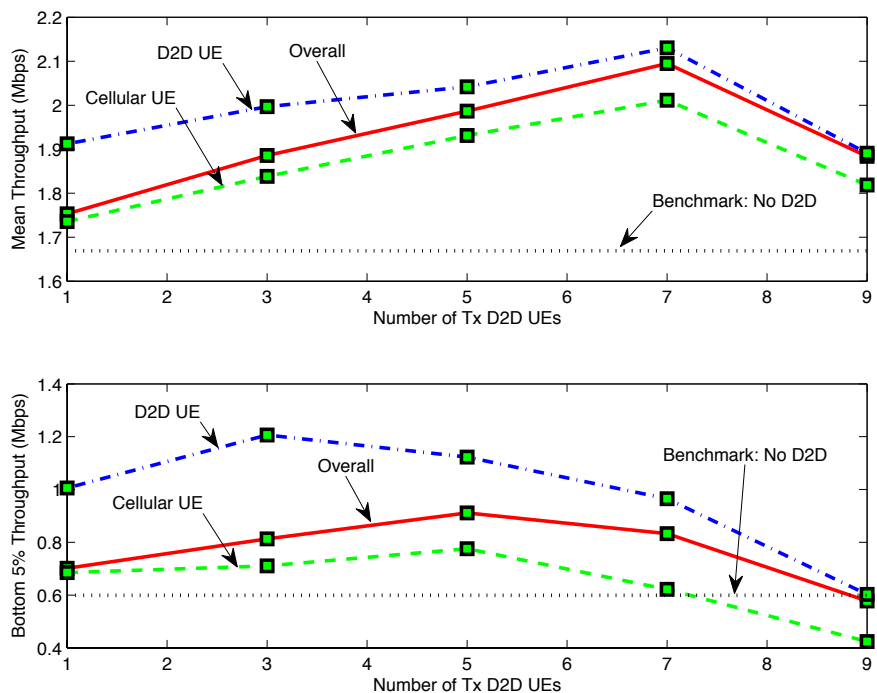

Fig. 4. Uplink throughput performance: 1 cell/3 sectors; ISD = 500 m; D2D range = 50 m; 10 transmitting UEs/sector (including both cellular UEs and transmitting D2D UEs); proportional fair scheduling is applied.

## V. Conclusions

D2D is an exciting and innovative feature that is very likely to be present in LTE after Release 12; it will facilitate the interoperability between critical public safety networks and ubiquitous commercial networks based on LTE. D2D fundamentally alters the cellular architecture, reducing the primacy of the BS and enabling UEs to transmit directly to nearby "friends". As this article discussed in detail, such a shift requires a rethinking of many of the working assumptions and models used to date for cellular systems. This article has particularly focused on current D2D standardization activities in 3GPP for LTE, although most of the conclusions herein would likely apply to any D2D-enhanced cellular standard.